\documentclass[12pt]{article}

\pdfoutput=1

\usepackage[bulletsep]{collref}
\usepackage{amssymb,graphicx}
\usepackage[intlimits]{amsmath}
\usepackage{bbm}
\usepackage[small]{subfigure}

\usepackage{MnSymbol}

\makeatletter \@addtoreset{equation}{section} \makeatother

\makeatletter
\let\old@startsection=\@startsection
\let\oldl@section=\l@section
\renewcommand{\@startsection}[6]{\old@startsection{#1}{#2}{#3}{#4}{#5}{#6\mathversion{bold}}}
\renewcommand{\l@section}[2]{\oldl@section{\mathversion{bold}#1}{#2}}
\makeatother

\makeatletter
\let\old@makecaption=\@makecaption
\def\@makecaption{\small\old@makecaption}
\makeatother

\renewcommand{\leq}{\leqslant}

\begin{document}

\newcommand{\be}{\begin{equation}}\newcommand{\ee}{\end{equation}}
\newcommand{\bea}{\begin{eqnarray}} \newcommand{\eea}{\end{eqnarray}}
\def\p{\partial}
\def\pa{\partial}
\def\ov{\over }
\def\a{\alpha }
\def\g{\gamma}
\def\s{\sigma }
\def\td{\tilde }
\def\vp{\varphi}
\def\gd{\nu }
\def \ha {{1 \over 2}}

\def\KK{{\cal K}}

\def\Xint#1{\mathchoice
{\XXint\displaystyle\textstyle{#1}}
{\XXint\textstyle\scriptstyle{#1}}
{\XXint\scriptstyle\scriptscriptstyle{#1}}
{\XXint\scriptscriptstyle\scriptscriptstyle{#1}}
\!\int}
\def\XXint#1#2#3{{\setbox0=\hbox{$#1{#2#3}{\int}$ }
\vcenter{\hbox{$#2#3$ }}\kern-.5\wd0}}
\def\ddashint{\Xint=}
\def\dashint{\Xint-}

\newcommand\cev[1]{\overleftarrow{#1}}

\begin{flushright}\footnotesize
\texttt{NORDITA-2013-97} \\
\texttt{UUITP-21/13}
\vspace{0.6cm}
\end{flushright}

\renewcommand{\thefootnote}{\fnsymbol{footnote}}
\setcounter{footnote}{1}

\begin{center}
{\Large\textbf{\mathversion{bold} Localization at Large $N$\footnote{Talk by K.Z. at "Pomeranchuk-100", Moscow, 5-6 June 2013. To be published in the proceedings.}}
\par}

\vspace{0.8cm}

\textrm{J.G.~Russo$^{1,2}$ and
K.~Zarembo$^{3,4,5}$}
\vspace{4mm}

\textit{${}^1$ Instituci\'o Catalana de Recerca i Estudis Avan\c cats (ICREA), \\
Pg. Lluis Companys, 23, 08010 Barcelona, Spain}\\
\textit{${}^2$  Department ECM, Institut de Ci\`encies del Cosmos,  \\
Universitat de Barcelona, Mart\'\i \ Franqu\`es, 1, 08028 Barcelona, Spain}\\
\textit{${}^3$Nordita, KTH Royal Institute of Technology and Stockholm University,
Roslagstullsbacken 23, SE-106 91 Stockholm, Sweden}\\
\textit{${}^4$Department of Physics and Astronomy, Uppsala University\\
SE-751 08 Uppsala, Sweden}\\
\textit{${}^5$Institute of Theoretical and Experimental Physics, B. Cheremushkinskaya 25, 117218 Moscow, Russia}\\

\vspace{0.2cm}
\texttt{jorge.russo@icrea.cat, zarembo@nordita.org}

\vspace{3mm}


\par\vspace{1cm}

\textbf{Abstract} \vspace{3mm}

\begin{minipage}{13cm}

We review how localization is used to probe holographic duality and, more generally, non-perturbative dynamics of four-dimensional $\mathcal{N}=2$ supersymmetric gauge theories in the planar large-$N$ limit.

\end{minipage}

\end{center}

\vspace{0.5cm}



\setcounter{page}{1}
\renewcommand{\thefootnote}{\arabic{footnote}}
\setcounter{footnote}{0}

\section{Introduction}

String theory on $AdS_5\times S^5$ gives a holographic description of the superconformal $\mathcal{N}=4$ Yang-Mills (SYM) through the AdS/CFT correspondence \cite{Maldacena:1998re,Gubser:1998bc,Witten:1998qj}. This description is exact, it maps correlation functions in SYM, at any coupling, to string amplitudes in $AdS_5\times S^5$. Gauge-string duality for less supersymmetric and non-conformal theories is at present less systematic, and is mostly restricted to the classical gravity approximation, which in the dual field theory corresponds to the extreme strong-coupling regime. For this reason, any direct comparison of holography with the underlying field theory requires non-perturbative input on the field-theory side.

 There are  no general methods, of course, but in the basic AdS/CFT context a variety of tools have been devised to gain insight into  the strong-coupling behavior of $\mathcal{N}=4$ SYM, notably by exploiting integrability of this theory in the planar limit \cite{Beisert:2010jr}. Another approach is based on supersymmetric localization \cite{Pestun:2007rz}. Applied to the $\mathcal{N}=4$ theory, localization provides direct dynamical tests of the AdS/CFT correspondence, but localization does not require as high supersymmetry as $\mathcal{N}=4$ and more importantly does not rely on conformal invariance thus allowing one to explore a larger set of models including massive theories. Some of localizable theories have known gravity duals, opening an avenue for direct comparison of holography with the first-principle field-theory calculations. In this contribution we concentrate on $\mathcal{N}=2$ theories in four dimensions. A  review of  localization in $D=3$ and its applications to the $AdS_4/CFT_3$ duality can be found in \cite{Marino:2011nm}; early results for $\mathcal{N}=4$ SYM \cite{Erickson:2000af,Drukker:2000rr} are reviewed in \cite{Semenoff:2002kk}.
 
 Although our main goal is holography and hence strong coupling, localization gives access to more general aspects of non-perturbative dynamics. An example of a non-perturbative phenomenon captured by localization is all-order OPE \cite{Russo:2013kea}.
 Suppose that we integrate out a heavy field of mass $M$  in an asymptotically free theory with a dynamically generated scale $\Lambda _{\rm eff}\ll M$. We then expect that any observable will  have an expansion 
\begin{equation}\label{genOPE}
 \mathcal{A}=\Lambda _{\rm eff}^\Delta \sum_{n=0}^{\infty }C_n\left(\frac{\Lambda _{\rm eff}}{M}\right)^{2n},
\end{equation}
 where $\Delta $ is the scaling dimension of $\mathcal{A}$. Expansions of this type make prominent appearance in the ITEP sum rules \cite{Shifman:1978bx,Shifman:1978by}. The mass $M$ in the denominator arises from expanding the effective action in local operators and powers of $\Lambda _{\rm eff}$ in the numerator come from the condensates, the vacuum expectation values of local operators generated by the OPE. The coefficients in this expansion carry non-perturbative information, and are usually difficult to calculate, but for observables amenable to localization it is possible to compute these coefficients to all orders.
 
 \begin{figure}[t]
\begin{center}
\subfigure[]{
   \includegraphics[height=3.5cm] {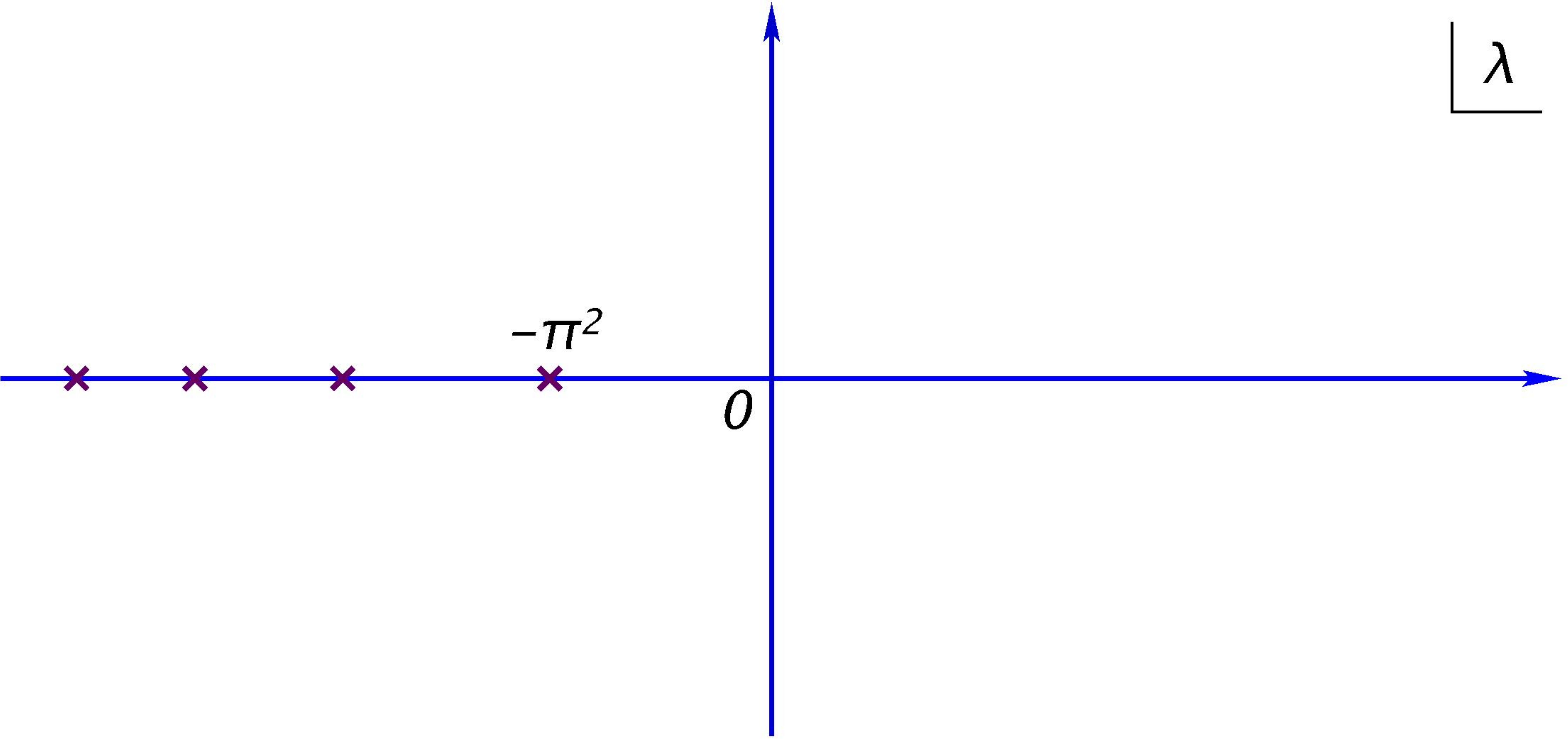}
   \label{fig:subfig1}
 }
 \subfigure[]{
   \includegraphics[height=3.5cm] {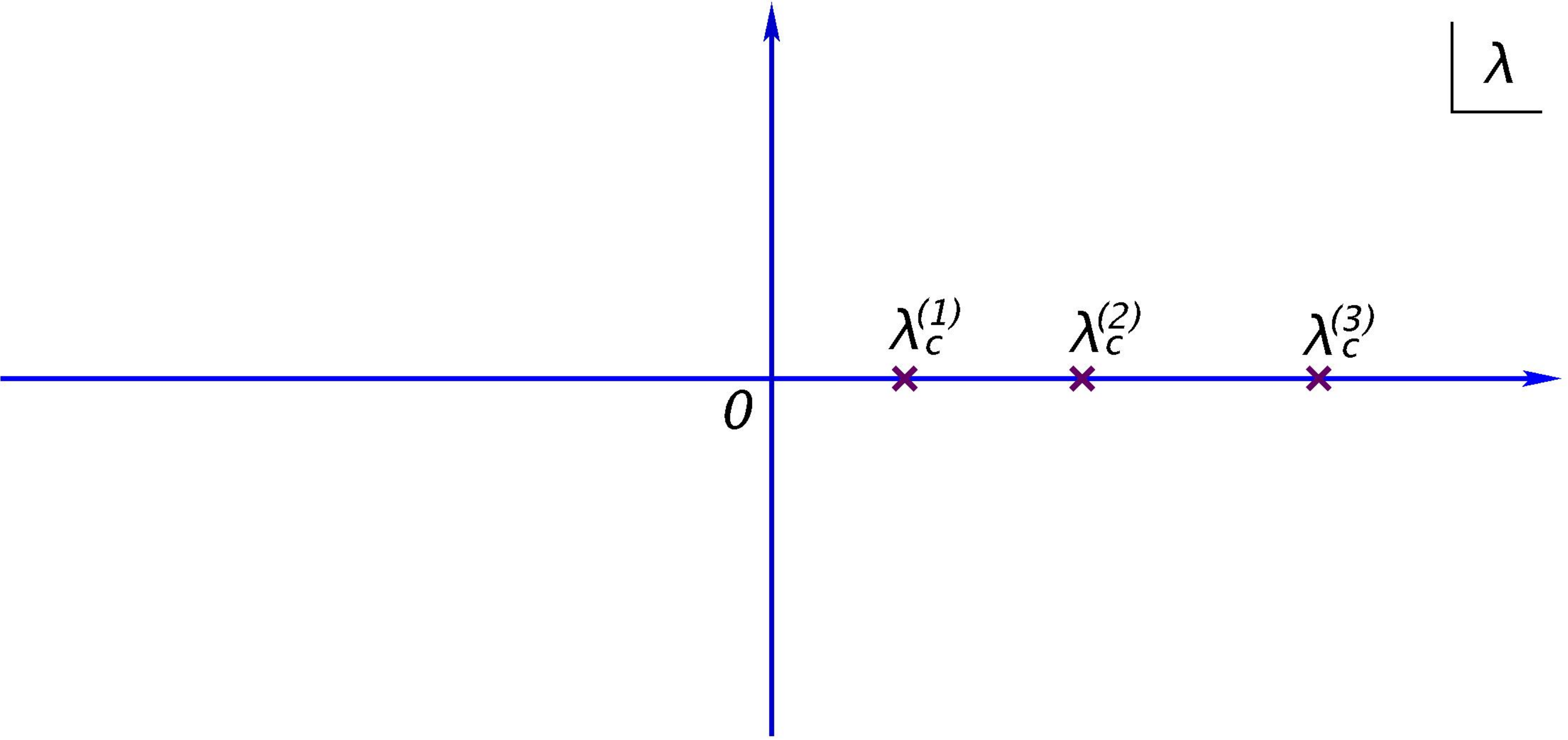}
   \label{fig:subfig2}
 }
\caption{\label{}\small Possible analytic structure in the coupling constant plane: (a) No phase transitions (all singularities lie on the negative part of the real axis). This case is realized in $\mathcal{N}=4$ SYM. (b) The singularities lie on the positive real axis, leading to phase transitions at $\lambda =\lambda _c^{(n)}$.}
\end{center}
\end{figure}
 
The OPE of the form (\ref{genOPE}) is of course expected on general grounds in the regime $\Lambda _{\rm eff}\ll M$. What is less expected, but appears to be generic, is  the emergence of large-$N$ phase transitions when $\Lambda _{\rm eff}\gtrsim M$ \cite{Russo:2013qaa,Russo:2013kea}. Large-$N$ phase transitions are very familiar from matrix models \cite{Gross:1980he,Wadia:2012fr} as singularities reflecting the finite radius of convergence of planar perturbation theory \cite{Brezin:1977sv}. For any UV finite theory, including $\mathcal{N}=4$ SYM, planar perturbation theory also has a finite radius of convergence. But in the $\mathcal{N}=4$ case all singularities lie on the negative real axis of the 't~Hooft coupling $g_{\rm YM}^2N\equiv \lambda$ (the leading singularity appears at $\lambda =-\pi ^2$). Interpolation from weak to strong coupling is thus continuous   as illustrated in fig.~\ref{fig:subfig1}. 
There are no distinct ``perturbative" and ``holographic" phases. {\it A priori} this does not follow from any fundamental principle. A possibility that  $\mathcal{N}=4$ SYM undergoes a strong-weak phase transition  was in fact contemplated in the early days of AdS/CFT \cite{Li:1998kd}, but subsequent developments showed that such a transition does not occur. Is it still possible that theories different from $\mathcal{N}=4$ SYM have a structure of singularities shown in fig.~\ref{fig:subfig2}? And, if yes, what are the implications for the holographic duality? Localization gives partial answers to these questions: phase transitions do occur in massive theories, but it is not clear at the moment how to describe them holographically.

\section{Localization in $\mathcal{N}=2^*$ SYM and large-$N$ limit}\label{sec:loc}

Our prime example will be the $\mathcal{N}=2^*$ theory, a massive deformation of $\mathcal{N}=4$ SYM which preserves half of the supersymmetry:
\begin{equation}\label{Lagrang-N2star}
 \mathcal{L}_{\mathcal{N}=2^*}=\mathcal{L}_{\mathcal{N}=4}
 +M\mathcal{O}_3+M^2\mathcal{O}_2.
\end{equation}
The dimension two and dimension three operators $\mathcal{O}_{2,3}$ give masses to four out of six adjoint scalars and to half of the fermions, and also contain certain tri-linear couplings. Two scalars that remain massless belong to the vector multiplet of $\mathcal{N}=2$ supersymmetry: $(A_\mu , \psi^1_\alpha  , \psi^2_\alpha  , \Phi, \Phi ')$, while massive fields combine to the complex hypermultiplet  $(\phi , \chi _\alpha ,\tilde{\chi }_\alpha ,\tilde{\phi })$, also  in the adjoint representation of the $SU(N)$ gauge group.  

This theory inherits finiteness of the $\mathcal{N}=4$ SYM. 
The holographic description of $\mathcal{N}=2^*$ SYM at strong coupling \cite{Pilch:2000ue,Buchel:2000cn} is based on the solution of type IIB supergravity in ten dimensions that was obtained by Pilch and Warner \cite{Pilch:2000ue} by perturbing $AdS_5\times S^5$  with constant sources dual to relevant operators in the Lagrangian (\ref{Lagrang-N2star}).

\begin{figure}[t]
\begin{center}
   \includegraphics[height=4cm] {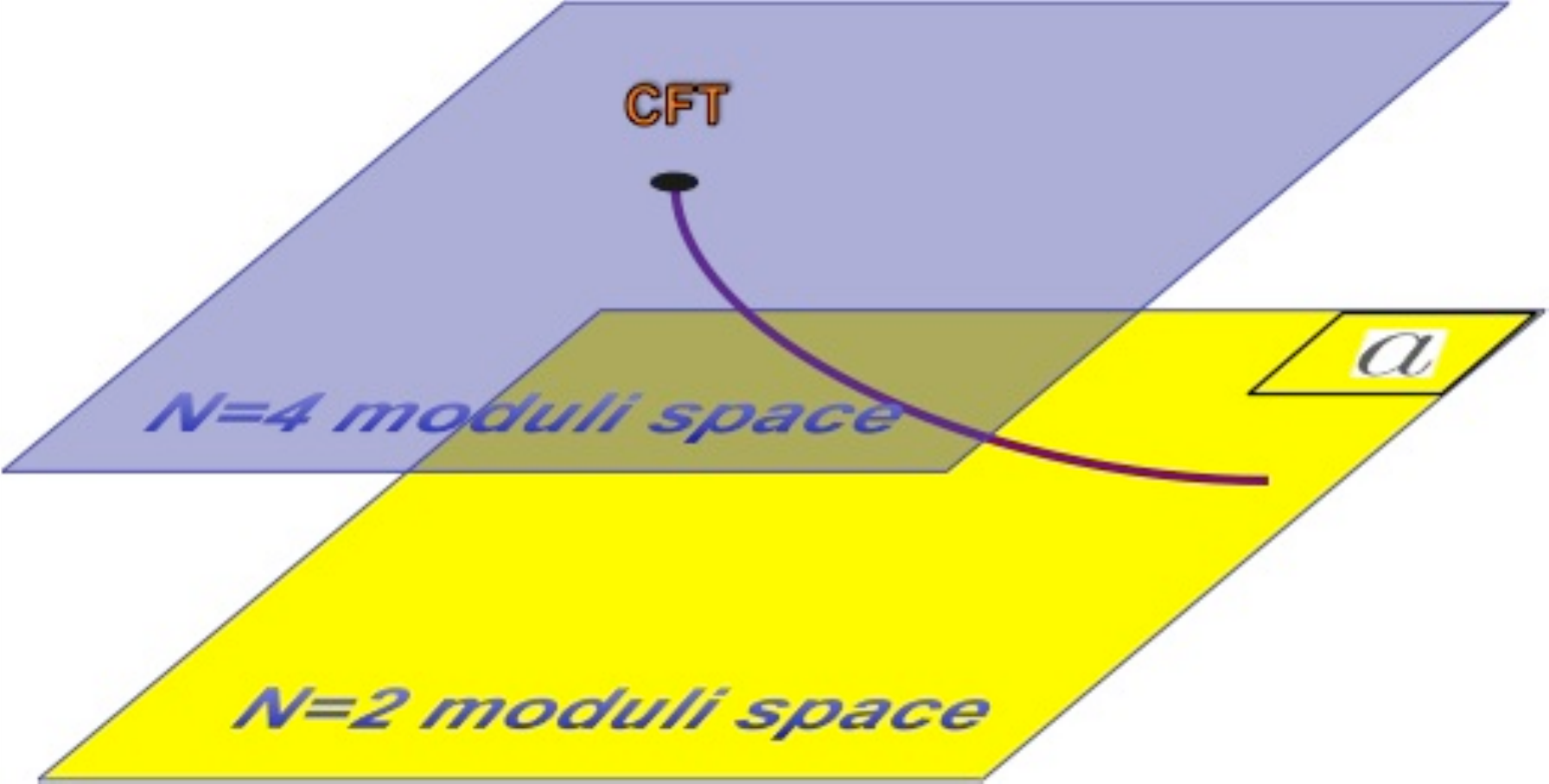}
\caption{\label{ModSpace}\small $\mathcal{N}=2^*$ flow trajectory in the moduli space of $\mathcal{N}=2$ theories.}
\end{center}
\end{figure}

The  moduli space of vacua of $\mathcal{N}=2^*$ SYM is parameterized by the diagonal expectation value of the adjoint scalar in the vector multiplet:
\begin{equation}\label{Phivev}
 \left\langle \Phi \right\rangle=\mathop{\mathrm{diag}}\left(a_1,\ldots ,a_N\right).
\end{equation}
The same vacuum degeneracy exists in $\mathcal{N}=4$ SYM as well, where any of the six scalars can take on a vev. However, when talking about $\mathcal{N}=4$ SYM, we will always assume that the theory is at the conformal point with zero vevs. We also assume that the $\mathcal{N}=2^*$ flow  starts with this conformal theory at the origin of the moduli space. In other words, by $\mathcal{N}=2^*$ SYM we really mean a line of theories that at $M=0$ degenerate to the conformal state of $\mathcal{N}=4$ SYM with $\left\langle \Phi \right\rangle=0$. The flow so defined  traces a trajectory in the moduli space of vacua as schematically illustrated in fig.~\ref{ModSpace}. 
At the IR end of the flow, hypermultiplets become very heavy and can be integrated out leaving behind pure $\mathcal{N}=2$ SYM with the dynamically generated  scale $\Lambda _{\rm eff}=M\exp(-4\pi ^2/\lambda )$. It is important to keep in mind that
the $(\mathcal{N}=4)\rightarrow (\mathcal{N}=2)$ flow picture is only correct at sufficiently weak coupling when the scales $M$ and $\Lambda _{\rm eff}$ are largely separated. For finite coupling, $\Lambda _{\rm eff}$ is of the form
$ \Lambda_{\rm eff} = g(\lambda) M$, with calculable $g(\lambda)$, which starts off exponentially small, but becomes much greater than one at $\lambda \to\infty$, where $ \Lambda_{\rm eff}$  exceeds $M$.

Normally the scalar vev (\ref{Phivev}) can be chosen at will, but in the flow that starts at the origin of the  moduli space, the vev is fixed by dynamics. Since in practice we will use a definition of the $\mathcal{N}=2^*$ SYM that is different from conformal perturbation theory, it is worthwhile to discuss the vacuum selection mechanism in some more detail. Consider, for the sake of illustration, a Heisenberg ferromagnet whose degenerate set of vacua is a two-sphere. A commonly used definition of the theory consists in slightly lifting the vacuum degeneracy, for instance by switching  on an external magnetic field, and then adiabatically relaxing the field to zero. The system will end up with magnetization aligned with the original direction of the field, or, depending on the temperature and details of the spin interactions, in the disordered state with no magnetization. These two regimes are separated by a phase transition.

The vacuum selection in the $\mathcal{N}=2^*$ theory can also be defined by switching on  infinitesimal external perturbation that lifts the vacuum degeneracy. A natural choice would be temperature, which generates a potential on the moduli space. The path integral  on $\mathbbm{R}^4$ would then be defined by taking the zero-temperature limit. Temperature here plays the same 
r\^ole as the magnetic field for the ferromagnet. 
While selecting thermalizable vacuum is physically appealing,  supersymmetry breaking makes this procedure difficult to implement in practice. An alternative
procedure, the one that we are going to use, consists in compactifying the theory on $S^4$. The curvature couplings generate a scalar potential which completely lifts the vacuum degeneracy, while the  path integral on $S^4$ includes integration over zero modes of all fields and is thus well defined without specifying any boundary conditions. Sending subsequently the radius of the sphere to infinity defines a unique vacuum in the decompactification limit. We conjecture that the $S^4$ vacuum actually coincides with the thermalizable state, and that both definitions  are equivalent to conformal perturbation theory in $\mathcal{N}=4$ SYM with zero scalar vev.  The advantage of the $S^4$ compactification is that the path integral on the sphere can be computed exactly using localization  \cite{Pestun:2007rz}. From this point of view the radius of the sphere is just an IR regulator, playing the same r\^ole as the magnetic field in the ferromagnet, but it is also of some interest to keep the radius finite introducing an extra parameter into the theory.  

At strong coupling all three ways to define the $\mathcal{N}=2^*$ flow are manifestly equivalent. Indeed, the Pilch-Warner solution is obtained by perturbing $AdS_5\times S^5$ with the hypermultiplet mass and thus by construction is dual to the flow trajectory that ends in the conformal point on the $\mathcal{N}=4$ moduli space (fig.~\ref{ModSpace}). On the other hand, the Pilch-Warner background can be  thermalized  \cite{Buchel:2003ah} as well as compactified on $S^4$ \cite{Bobev:2013cja}. The  latter case can be directly compared to field theory through localization  \cite{Buchel:2013id,Bobev:2013cja} thus providing further  evidence that the $S^4$ vacuum is dual to the Pilch-Warner solution at strong coupling.

In the planar limit the vacuum is characterized by the master field, a distribution of the eigenvalues of the adjoint scalar from the vector multiplet:
\be\label{eigendensity}
\rho(x) = \left\langle\frac{1}{N} \sum_{i=1}^N \delta(x-a_i)\right\rangle.
\ee
 Using the localization results of \cite{Pestun:2007rz}, we can explicitly calculate the large-$N$ master field on a sphere of any radius. The $S^4$ vacuum is obtained by sending the radius to infinity. However, keeping the radius finite is interesting in its own right, and we will also discuss dependence of the master field on the radius of compactification.

In theories with a fermionic symmetry the exact functional integral may localize in a subset of field
configurations  plus a one-loop contribution. The $\mathcal{N}=2^*$ theory belongs to this class, as its partition function on $S^4$ reduces to a finite-dimensional integral over   the eigenvalues of the adjoint scalar in the vector multiplet (\ref{Phivev}). 
The result can be expressed as follows \cite{Pestun:2007rz}:
\be\label{mint}
Z_{{\cal N}=2^*}=\int \prod_{i}^{}da_i\,\delta \left(\sum_{i}^{}a_i\right)
\prod _{i<j}\frac{(a_i-a_j)^2H^2(a_i-a_j)}{H(a_i-a_j-M)H(a_i-a_j+M)}
\,{\rm e}\,^{-\frac{8\pi^2N}{\lambda}\sum_i a_i^2 } \left|\mathcal{Z}_{\rm inst}\right|^2,
\ee
where $H(x)$ encodes the product over the spherical harmonics of all field fluctuations:
\begin{equation}\label{functionH}
 H(x)\equiv \prod_{n=1}^\infty \left(1+\frac{x^2}{n^2}\right)^n \,{\rm e}\,^{-\frac{x^2}{n}} .
\end{equation}
 The one-loop contribution  of the vector  multiplet combines into $H^2$ in the numerator, while that of the hypermultiplet gives the two $H$'s in the denominator. 
The instanton factor $\mathcal{Z}_{\rm inst}$ is the Nekrasov partition function \cite{Nekrasov:2002qd,Nekrasov:2003rj,Okuda:2010ke} with the equivariant parameters given by $\epsilon _1=\epsilon _2=1$. 
In the large-$N$ limit the instantons are suppressed by a factor\footnote{
The volume of the instanton moduli space could potentially compensate the exponential suppression \cite{Gross:1994mr}, but explicit evaluation of the one-instanton weight at large $N$ indicates that this never happens in the $\mathcal{N}=2^*$ theory, the instantons always remaining suppressed at
 $N\rightarrow \infty $ \cite{Russo:2013kea}.
And indeed  the results of localization with instantons neglected are in perfect agreement with the supergravity predictions  at strong coupling \cite{Buchel:2013id,Bobev:2013cja}.} $\exp(-8\pi^2N|k|/\lambda)$,
where $k$ is the instanton number. Since we are interested in the planar limit, we will just set $\mathcal{Z}_{\rm inst}=1$.

For notational convenience the radius of the sphere has been set to one. The dependence on $R$ is reinstated
by rescaling $a_i\rightarrow a_iR$ and $M\rightarrow MR$. We mostly use dimensionless units throughout the paper, but will recover the dependence on $R$ when discussing the decompactification limit  $R\rightarrow \infty $.
In the dimensionless units, decompactification is equivalent to the infinite-mass limit $M\rightarrow \infty $. 

The eigenvalue integral (\ref{mint}) is (literally!) infinitely simpler than the path integral of quantum field theory, making localization a very powerful computational tool, especially at large $N$ when instantons can be neglected and the eigenvalue integral can be analyzed by methods familiar from random matrix theory \cite{Brezin:1977sv}. This simplicity comes at a price of dealing with a limited number of observables. One quantity that can be calculated with the help of localization is 
the free energy
\begin{equation}\label{freenegry-def}
 F=-\frac{1}{N^2}\,\ln Z.
\end{equation}
 Another one is expectation value of the circular Wilson loop. This couples to the gauge field and the scalar from the vector multiplet as follows:
\begin{equation}\label{WilsonLoop}
 W(C)\equiv \left\langle \frac{1}{N}\,\mathop{\mathrm{tr}}{\rm P}\exp
 \left[\oint_C d\tau \,\left(i\dot{x}^\mu A_\mu + |\dot{x}|\Phi \right)\right]
 \right\rangle.
\end{equation}
If the contour $C$ goes around the equator of the four-sphere, the fields can be replaced by their classical values, $A_\mu =0$ and $\Phi $ given by (\ref{Phivev}). Therefore
\begin{equation}
 \label{wils}
W(C)=\left\langle \frac{1}{N}\sum_{i}^{}\,{\rm e}\,^{2\pi a_i}\right\rangle.
 \end{equation}
For the circular Wilson loop the expectation value maps to the average of the exponential operator in the matrix model
 (\ref{mint}).
 
 In the large $N$ limit, the eigenvalue integral (\ref{mint}) is of the saddle-point type. The saddle-point equations are the force balance conditions for  $N$ particles on a line with pairwise interactions which are subject to a common external potential:
 \be
\sum_{j\neq i}\left( \frac{1}{a_i-a_j} 
 -\mathcal{K}(a_i-a_j)
 +\frac{1}{2}\,\mathcal{K}(a_i-a_j+M)
 +\frac{1}{2}\,\mathcal{K}(a_i-a_j-M)
\right)=\frac{8\pi^2 N}{\lambda}\, a_i.
\ee
where 
\begin{equation}\label{functionKK}
 \KK(x) \equiv  -\frac{H'(x)}{H(x)} =2x\sum_{n=1}^\infty \left(\frac{1}{n} -\frac{n}{n^2+x^2}\right).
\end{equation}
These equations are  equivalent to a singular integral equation for the eigenvalue density (\ref{eigendensity}):
\begin{equation}
 \label{nnstar}
\strokedint_{-\mu}^\mu dy\, \rho(y)
\left(\frac{1}{x-y} -\KK(x-y)+\frac{1}{2}\,\KK(x-y+M)+\frac{1}{2}\,\KK(x-y-M)\right)= \frac{8\pi^2}{\lambda}\ x,\qquad 
x\in [-\mu,\mu ].
\end{equation}
Here we have assumed that eigenvalues are distributed in one cut along the interval $[-\mu,\mu ]$, where
the density $\rho (x)$ is unit-normalized.

 Once the integral equation is solved, the Wilson loop can be computed from the Laplace transform of the density:
\begin{equation}
 W(C)=\int_{-\mu }^{\mu }dx\,\rho (x)\,{\rm e}\,^{2\pi x}\equiv \left\langle \,{\rm e}\,^{2\pi x}\right\rangle.
\end{equation}
The free energy is given by a double integral. It is actually easier to compute its derivatives, for instance,
\begin{equation}\label{heatcap}
 \frac{\partial F}{\partial \lambda }=-\frac{8\pi ^2}{\lambda ^2}\,\left\langle x^2\right\rangle.
\end{equation}

\section{Strong coupling and holography}

Having the exact coupling dependence for the planar theory, one can study the important limit $\lambda \gg 1$, which explores the deep quantum regime of the theory.
An extra motivation for studying
the $\lambda \gg 1$ limit is that this  is precisely the limit where  super Yang-Mills theories are expected to have a
holographic description in terms of a weakly-curved supergravity dual.
In this section we shall discuss two examples: the superconformal $\mathcal{N}=4$ SYM and its mass deformation preserving  $\mathcal{N}=2$
supersymmetry.

\subsection {$\mathcal{N}=4$ SYM}

Since $\mathcal{N}=4$ SYM is a conformal theory and the sphere  is conformally equivalent to $\mathbbm{R}^4$, one can use localization to compute a circular Wilson loop in flat space. The answer does not depend on the radius of the circle by conformal invariance, and is given by the exponential average in the Gaussian matrix model \cite{Erickson:2000af,Drukker:2000rr,Pestun:2007rz}:
\begin{equation}\label{randmatN4}
 W(C_{\rm ircle})=\left\langle \frac{1}{N}\,\mathop{\mathrm{tr}}\,{\rm e}\,^{2\pi \Phi }\right\rangle,
 \qquad 
 Z=\int_{}^{}d\Phi \,\,{\rm e}\,^{-\frac{8\pi ^2N}{\lambda }\,\mathop{\mathrm{tr}}\Phi ^2}.
\end{equation}
This is equivalent to (\ref{mint}) with $M=0$, upon gauge fixing the matrix measure to eigenvalues. 

\begin{figure}[t]
\begin{center}
 \centerline{\includegraphics[width=10cm]{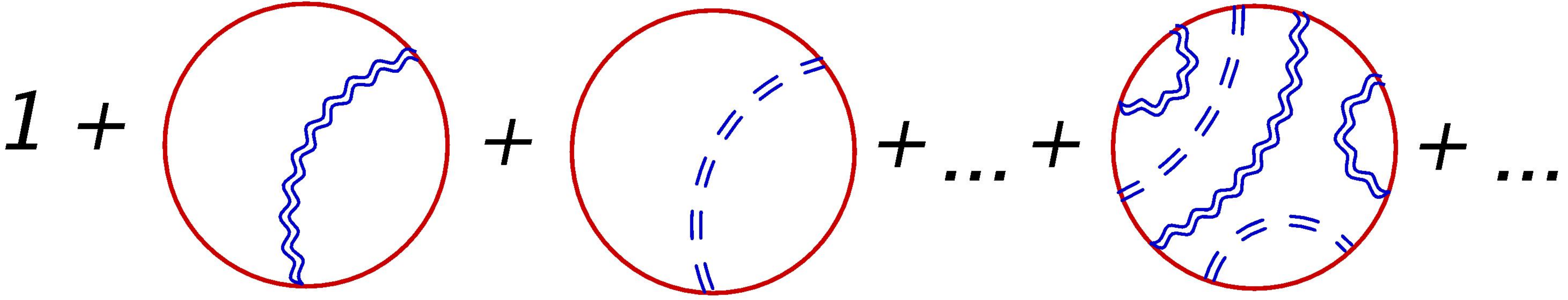}}
\caption{\label{rainbows}\small Rainbow diagrams for the circular Wilson loop.}
\end{center}
\end{figure}

In the localization approach, the random matrix $\Phi $ is the zero mode of the adjoint scalar on $S^4$. By construction it has a constant propagator. If we start directly from field theory on $\mathbbm{R}^4$, the result (\ref{randmatN4}) can be understood as resummation of rainbow diagrams -- all possible diagrams without internal vertices (fig.~\ref{rainbows}) \cite{Erickson:2000af,Drukker:2000rr}. 
One can argue that other diagrams do not contribute to the circular loop average \cite{Drukker:2000rr}. The constant propagator then arises from partial cancellation between the scalar and gluon exchanges. The numerator in the one-loop correction (the first two diagrams in fig.~\ref{rainbows}) contains the gauge boson $\dot{x}_1\cdot \dot{x}_2$ and scalar $|\dot{x}_1|\,|\dot{x}_2|$ contributions. For the circular loop, they combine into $|\dot{x}_1|\,|\dot{x}_2|-\dot{x}_1\cdot \dot{x}_2=1-x_1\cdot x_2$ and cancel the denominator $(x_1-x_2)^2=2-2x_1\cdot x_2$, leaving behind a constant propagator $\lambda /8\pi ^2$.  This argument extends to rainbow graphs of any order in perturbation theory, and the problem effectively reduces to combinatorics, taken into account by the matrix integral. There is no Feynman-diagram derivation for more complicated localization matrix models, but in the case of $\mathcal{N}=2$ superconformal QCD one can check that the first vertex correction that appears at three loops  \cite{Andree:2010na} can be consistently reproduced from the skeleton graphs of the zero-dimensional matrix integral \cite{Passerini:2011fe}.

For later reference we do a simple exercise of solving the Gaussian model (\ref{randmatN4})  at large $N$.  The saddle-point equation (\ref{nnstar}) at $M=0$ becomes 
\begin{equation}
 \label{nn4}
\strokedint_{-\mu}^\mu dy\, \,
\frac{\rho(y)}{x-y} = \frac{8\pi^2}{\lambda}\ x.
\end{equation}
The eigenvalue density is then given by the Wigner's semi-circle law:
 \be
 \label{wig}
 \rho(x)= \frac{8\pi}{\lambda} \sqrt{ \frac{\lambda}{4\pi^2} -x^2},
 \ee
 and from (\ref{wils}) we get the vacuum expectation value of the circular Wilson loop \cite{Erickson:2000af}:
\be
W(C_{\rm circle})= \int dx\, \rho(x)\ {\rm e}^{2\pi x}=\frac{2}{\sqrt{\lambda}}\, I_1\left(\sqrt{\lambda}\right).
\ee
The free energy can be inferred from (\ref{heatcap}), or calculated directly by doing  the Gaussian integral in (\ref{randmatN4}):
\be\label{Fmmod}
 F=-\frac{1}{2}\,\ln\lambda.
\ee
In the strong coupling $\lambda\gg 1 $ regime, the Wilson loop has the behavior
\be\label{lambda34}
W \simeq \sqrt{\frac{2}{\pi}}\, \lambda^{-\frac{3}{4}}\, {\rm e}^{\sqrt{\lambda}}\ \qquad \left(\lambda\gg 1\right).
\ee
These results, obtained by directly computing the path integral in $\mathcal{N}=4$ SYM (or summing up Feynman diagrams) can be compared to the holographic predictions of string theory in $AdS_5\times S^5$.

\begin{figure}[t]
\begin{center}
 \subfigure[]{
   \includegraphics[height=4.7cm, trim=4cm 0cm 0cm 0cm] {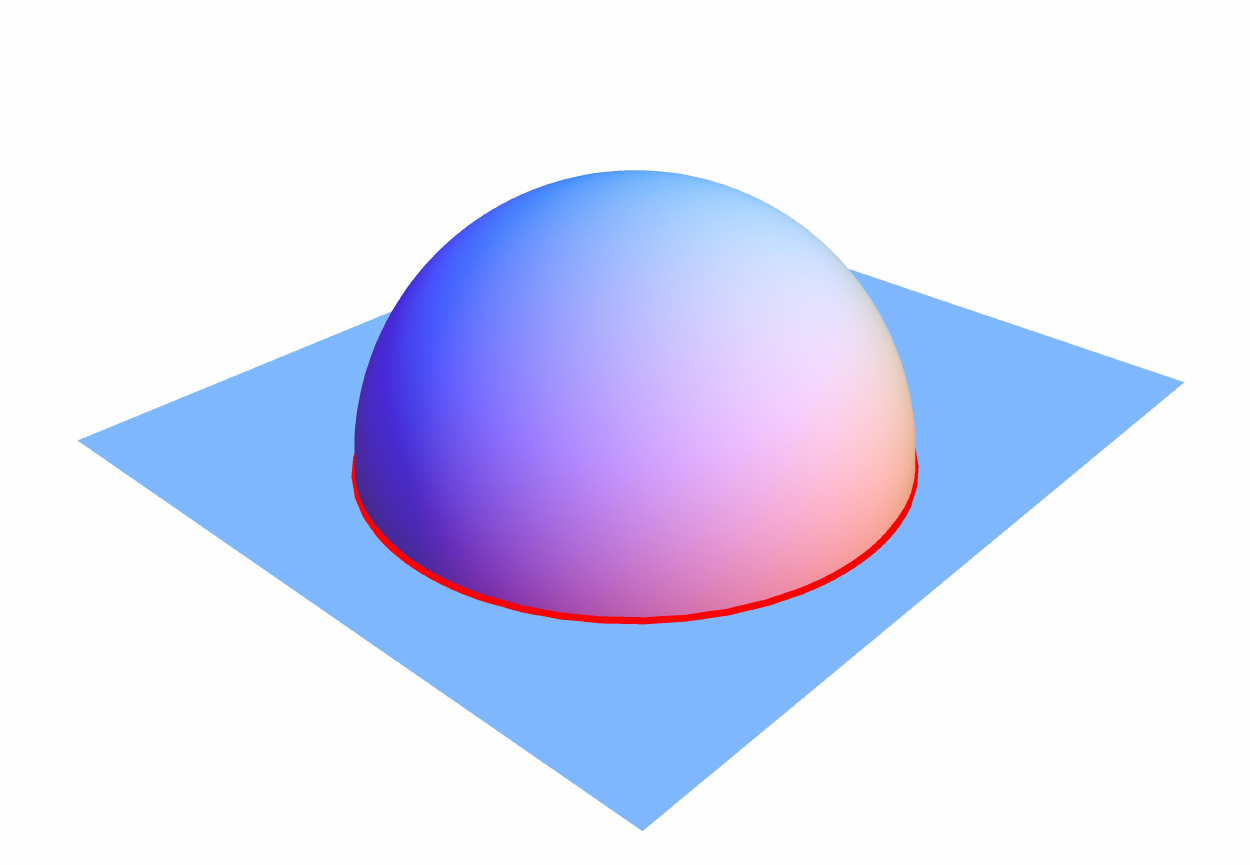}
   \label{fig2:subfig1}
 }
 \subfigure[]{
   \includegraphics[height=4.8cm, trim=1.7cm 2.5cm 0cm 0cm] {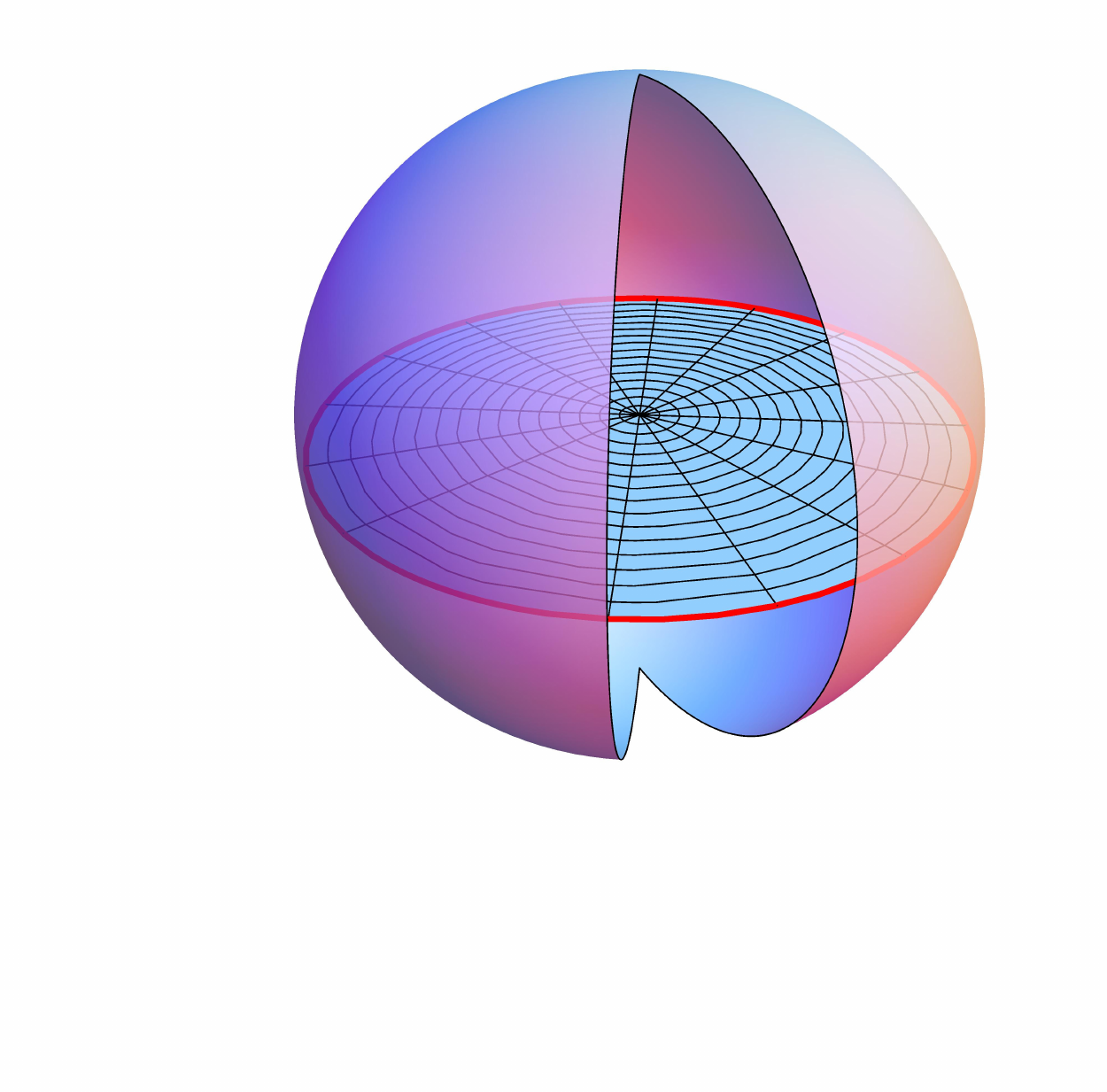}
   \label{fig2:subfig2}
 }
\caption{\label{}\small Minimal surface in $AdS_5$ that describes the circular Wilson loop: (a) in Poincar\'e coordinates; (b) in the $S^4$ slicing.}
\end{center}
\end{figure}

According to the AdS/CFT dictionary,  Wilson loop expectation values obey the area law  at strong coupling \cite{Maldacena:1998im,Rey:1998ik}:
\begin{equation}\label{GeneralAreaLaw}
 W(C)\simeq \,{\rm e}\,^{-\frac{\sqrt{\lambda }}{2\pi }\,A_{\rm reg}(C)},
\end{equation}
where $A_{\rm reg}(C)$ is the regularized area of the minimal surface  in $AdS_5$ that ends on the contour $C$, placed at the boundary, as illustrated in  fig.~\ref{fig2:subfig1}. The minimal surface ending on a circular contour was constructed in \cite{Drukker:1999zq,Berenstein:1998ij} in the Poincar\'e coordinates. For computing its area it is easier though to deal with the $S^4$ slicing of $AdS_5$:
\begin{equation}\label{AdS-metric}
 ds^2_{AdS_5}=\frac{dr^2}{1+r^2}+r^2d\Omega _{S^4}^2.
\end{equation}
If the Wilson loop is chosen to run along the big circle of $S^4$, as appropriate for comparing to localization (but the answer will be the same for any circular loop, by conformal invariance), the minimal surface will coincide with the equatorial plane, fig.~\ref{fig2:subfig2}. The area can be readily computed from (\ref{AdS-metric}):
\begin{equation}\label{Acir}
 A(C_{\rm circle})=2\pi \int_{0}^{\Lambda _0}\frac{dr\,r}{\sqrt{1+r^2}}
 =2\pi \Lambda _0-2\pi,
\end{equation}
where $\Lambda_0\rightarrow \infty  $ is a UV regulator.
The first, linearly divergent term should be removed by regularization. We thus get for the Wilson loop expectation value:
\begin{equation}
 W(C_{\rm circle})\simeq \,{\rm e}\,^{\sqrt{\lambda }},
\end{equation}
in agreement with the direct field-theory calculation \cite{Erickson:2000af,Drukker:2000rr}. 

The area law arises in the leading, semiclassical order of the strong-coupling expansion and gets corrections once the string fluctuations are taken into account. Since the disc partition function contains a factor of $\alpha '{}^{3/2}$ from gauge fixing the residual $SL(2,\mathbbm{R})$ conformal symmetry, the prefactor of the Wilson loop expectation value should  be proportional to $\lambda ^{-3/4}\sim \alpha '{}^{3/2}$  \cite{Drukker:2000rr}, which is indeed the case for (\ref{lambda34}).  The numerical coefficient in (\ref{lambda34}) is  the one-loop contribution of string fluctuations, also potentially calculable from string theory 
\cite{Kruczenski:2008zk,Kristjansen:2012nz}.

The holographic free energy of $\mathcal{N}=4$ SYM is given by the on-shell gravitational action,
\begin{equation}\label{SUGRA-action}
 F=-\frac{1}{8\pi^2 }\int_{AdS_5}^{}d^5x\,\sqrt{g}\left(\mathcal{R}+{12}\right),
\end{equation}
evaluated on the metric (\ref{AdS-metric}).  We have taken into account here that the five-dimensional Newton's constant in the dimensionless units that we use is given by $G_N=\pi/2N^2$. The factor of $N^2$ is already included in the definition of the free energy in (\ref{freenegry-def}). The substitution of (\ref{AdS-metric}) into (\ref{SUGRA-action}) results in a badly divergent integral:
\begin{equation}\label{N=4free-e}
 F=\frac{8}{3}\int_{0}^{\Lambda_0 }\frac{dr\,r^4}{\sqrt{1+r^2}}
 =\frac{2}{3}\,\Lambda _0^4-\frac{2}{3}\,\Lambda ^2_0
 +\frac{1}{2}\,\ln\Lambda_0 -\frac{7}{12}\,.
 \end{equation}
The gravitational action therefore has to be regularized by adding boundary counterterms \cite{Skenderis:2002wp}, much like in the calculation of the Wilson loop. There is one crucial difference though. The gravitational action, in contradistinction to the minimal area, requires a log-divergent counterterm.  The log has to be treated with care, as the answer will depend on the precise definition of the UV cutoff. The free energy, computed directly in field theory on $S^4$, is log-divergent too \cite{Burgess:1999vb},
and to match the two expressions it is necessary to use the same subtraction scheme in both cases. The radial coordinate of $AdS_5$ differs from the energy  scale of $\mathcal{N}=4$ SYM by a factor of $\sqrt{\lambda }$, and so  the UV cutoffs do \cite{Peet:1998wn,Bianchi:2001de}. To see this, one can compare the divergent part of the string action, given by the first term in (\ref{Acir}) multiplied by $\sqrt{\lambda }/2\pi $, with the action of a heavy probe in the $\mathcal{N}=4$ theory, equal to  $2\pi \Lambda $, where we identified the mass of the probe with the energy cutoff. Equating the two expressions we find that $\Lambda _0= 2\pi \Lambda /\sqrt{\lambda }$. In the field-theory regularization scheme, one subtracts $\ln \Lambda $ from the free energy (this subtraction is implicit in the matrix model). We thus conclude that the AdS/CFT prediction for the free energy, regularized in the field-theory scheme, is \cite{Russo:2012ay}
\begin{equation}
 F=-\frac{1}{2}\,\ln{\lambda }+\,{\rm const}\,,
\end{equation}
 which coincides with the matrix-model prediction (\ref{Fmmod}) up to an unimportant constant shift. For the Wilson loop, the string calculation gives the leading order of the strong-coupling expansion. The $\alpha  '$-corrections to the leading order result organize into an (asymptotic) power series in $1/\sqrt{\lambda }$. Interestingly, the gravity prediction for the free energy is exact, there are no $\alpha '$ corrections.

\subsection{$\mathcal{N}=2^*$ theory}

The integral equation (\ref{nnstar}) has a complicated kernel and, at present, cannot be solved analytically in the closed form. Following the work on related $\mathcal{N}=2$ theories \cite{Rey:2010ry,Passerini:2011fe,Bourgine:2011ie,Fraser:2011qa,Russo:2012ay}, approximate solutions have been constructed in various regimes in \cite{Russo:2012kj,Buchel:2013id,Russo:2013qaa,Russo:2013kea}.
The asymptotic large-$\lambda $ solution \cite{Buchel:2013id} is particularly simple. In writing down (\ref{nnstar}) we have assumed that eigenvalues are distributed in one cut along the interval $[-\mu,\mu ]$, where
the density $\rho (x)$ is unit-normalized.
The  one-cut solution  can be justified by starting at very weak coupling and gradually increasing $\lambda $.
The linear force $8\pi^2x/\lambda $ in the saddle-point equation (\ref{nnstar}) is attractive and pushes the eigenvalues towards the origin.
When $\lambda $ is small, this force is very strong and eigenvalues are distributed in a small interval.
As $\lambda $ is gradually increased, the linear force $8\pi^2x/\lambda $ becomes weaker, and the eigenvalue distribution expands to larger  intervals.
For a sufficiently large $\lambda $, the width of the eigenvalue distribution becomes much larger than the bare mass  of the $\mathcal{N}=2^*$ theory\footnote{Notice that $\mu $ plays the r\^ole of the effective IR scale of the theory.}: $\mu \gg M$. In this limit, most of the eigenvalues will satisfy $|x-y|\gg M$, $|x-y|\gg 1$, which justifies the following approximation:
\begin{equation}
 \frac{1}{2}\,\KK(x-y+M)+\frac{1}{2}\,\KK(x-y-M)-\KK(x-y)\approx
 \frac{1}{2}\,\KK''(x-y)M^2\approx \frac{M^2}{x-y}\, ,
\end{equation}
where we have used the asymptotic formula for $\KK(x)$,
\begin{equation}
\label{Klog}
\KK(x ) = x\ln x^2+2\gamma x+O(x^{-1}).
\ee
As a result, the net  effect of the complicated terms in the saddle-point equation reduces to  multiplicative renormalization of the Hilbert kernel:
\begin{equation}
 \frac{1}{x-y}~\longrightarrow~ \frac{1+M^2}{x-y}\ .
\end{equation}
The solution to the saddle-point equation  is again  the Wigner' semicircle 
\be\label{WignerLaw}
\rho(x)= \frac{2}{\pi\mu^2}\,\sqrt{\mu^2-x^2}\,,
\ee
but now with
\begin{equation}
\label{strongmu}
 \mu =\frac{\sqrt{\lambda \left(1+M^2\right)}}{2\pi}\,.
\end{equation}

For the Wilson loop this result implies the same behavior as in $\mathcal{N}=4$ SYM,  with $\lambda $ rescaled by $1+M^2$:
\begin{equation}\label{W(C)atlargelambda}
 \ln W(C)\simeq  \sqrt{\lambda \left(1+M^2\right)}.
\end{equation}
As far as the free energy is concerned, the dependence on $M$ is more complicated, and cannot be inferred just from (\ref{heatcap}). A more accurate calculation that keeps track of the $\lambda $-independent constant gives \cite{Buchel:2013id}:
\begin{equation}\label{Fatlargel}
 F=-\frac{1+M^2}{2}\,\ln\frac{\lambda \left(1+M^2\right)\,{\rm e}\,^{2\gamma +\frac{1}{2}}}{16\pi ^2}+\gamma +\frac{1}{4}-\ln 4\pi .
\end{equation}

The free energy can be compared with the gravitational action of the solution dual to ${\cal N} = 2^*$ theory on $S^4$. Such solution was recently constructed  \cite{Bobev:2013cja}, and its action perfectly matches the scheme-independent part of the matrix model prediction\footnote{The constant term  is chosen here to match the $\mathcal{N}=4$ result (\ref{Fmmod}) at $M=0$, but otherwise it is obviously scheme-dependent. The term proportional to $M^2$ also depends on regularization, since the free energy  is log-divergent. We follow the regularization scheme used by Pestun in deriving the matrix model from  the path integral \cite{Pestun:2007rz}. It is unclear to us how to implement precisely this scheme in the supergravity calculation. A pragmatic point of view, taken in \cite{Bobev:2013cja}, consists in comparing the third derivatives of the free energy to remove scheme-dependent ambiguities altogether.} after  implementing holographic renormalization to cancel the UV divergences, similar to those that appear in (\ref{N=4free-e}). To compare the Wilson loop, the solution of the five-dimensional supergravity obtained in \cite{Bobev:2013cja} has to be uplifted in ten dimensions, which has not been done so far.

However, we can compare the matrix model prediction (\ref{W(C)atlargelambda}) with generic expectations on the field theory in flat space, for which the supergravity dual  is known in the full ten-dimensional form \cite{Pilch:2000ue}. The flat-space limit can be reached by restoring the dependence on the radius of the four-sphere: $M\rightarrow MR$, $\mu \rightarrow \mu R$, and subsequently taking $R\rightarrow \infty $. The circular Wilson loop behaves as $\sqrt{\lambda }MR$  in this limit, and although we cannot compute Wilson loops for any other contour  using localization, it is quite natural to assert that any sufficiently big Wilson loop in $\mathcal{N}=2^*$ SYM obeys the perimeter law:
\begin{equation}\label{perlaw}
 W(C)\simeq \,{\rm e}\,^{\frac{\sqrt{\lambda }ML}{2\pi }},
\end{equation}
where $L$ is the length of the contour $C$. The coefficient $\sqrt{\lambda }M/2\pi $ is chosen to match localization prediction for the circle, and can be interpreted as finite mass renormalization of a heavy external probe. This result is expected to hold for any loop on $\mathbbm{R}^4$ and can be compared to the minimal area law in the Pilch-Warner geometry \cite{Buchel:2013id}.

The Pilch-Warner solution \cite{Pilch:2000ue} asymptotes to $AdS_5\times S^5$ near the boundary.  The mass scale $M$  is set geometrically by a domain wall placed at the distance $M$ along the radial direction. Beyond the domain wall the metric deviates substantially from that of $AdS_5$. The slice of the Pilch-Warner geometry, that is necessary for computation of the Wilson loop (\ref{WilsonLoop}),  has the following metric in the string frame:
\begin{equation}\label{PW-metr}
 ds=\frac{BM^2dx_\mu ^2}{c^2-1}+\frac{dc^2}{B\left(c^2-1\right)^2}\,,
\end{equation}
where
\begin{equation}
 B=c+\frac{c^2-1}{2}\,\ln\frac{c-1}{c+1}\,.
\end{equation}
This metric has the same asymptotics at $c\rightarrow 1$ as (\ref{AdS-metric}) has at $r\rightarrow \infty $, upon the coordinate transformation\footnote{The Pilch-Warner solution should be more appropriately compared to the slicing of $AdS_5$ in which the boundary is flat $\mathbbm{R}^4$. Since we only look at
$R=\infty $ limit, the difference is immaterial here.}
\begin{equation}\label{cr-}
 c=1+\frac{M^2}{2r^2}\,.
\end{equation}
The domain wall is located at $c\sim 1$, and thus $r\sim M$. Beyond the domain wall the metric behaves as
\begin{equation}\label{farbeyond}
 ds^2\simeq \frac{1}{c^3}\left(\frac{2}{3}\,M^2dx_\mu ^2+\frac{3}
 {2}\,dc^2\right)
 \qquad \left(c\gg 1\right).
\end{equation}

\begin{figure}[t]
\begin{center}
\subfigure[]{
   \includegraphics[height=4.5cm] {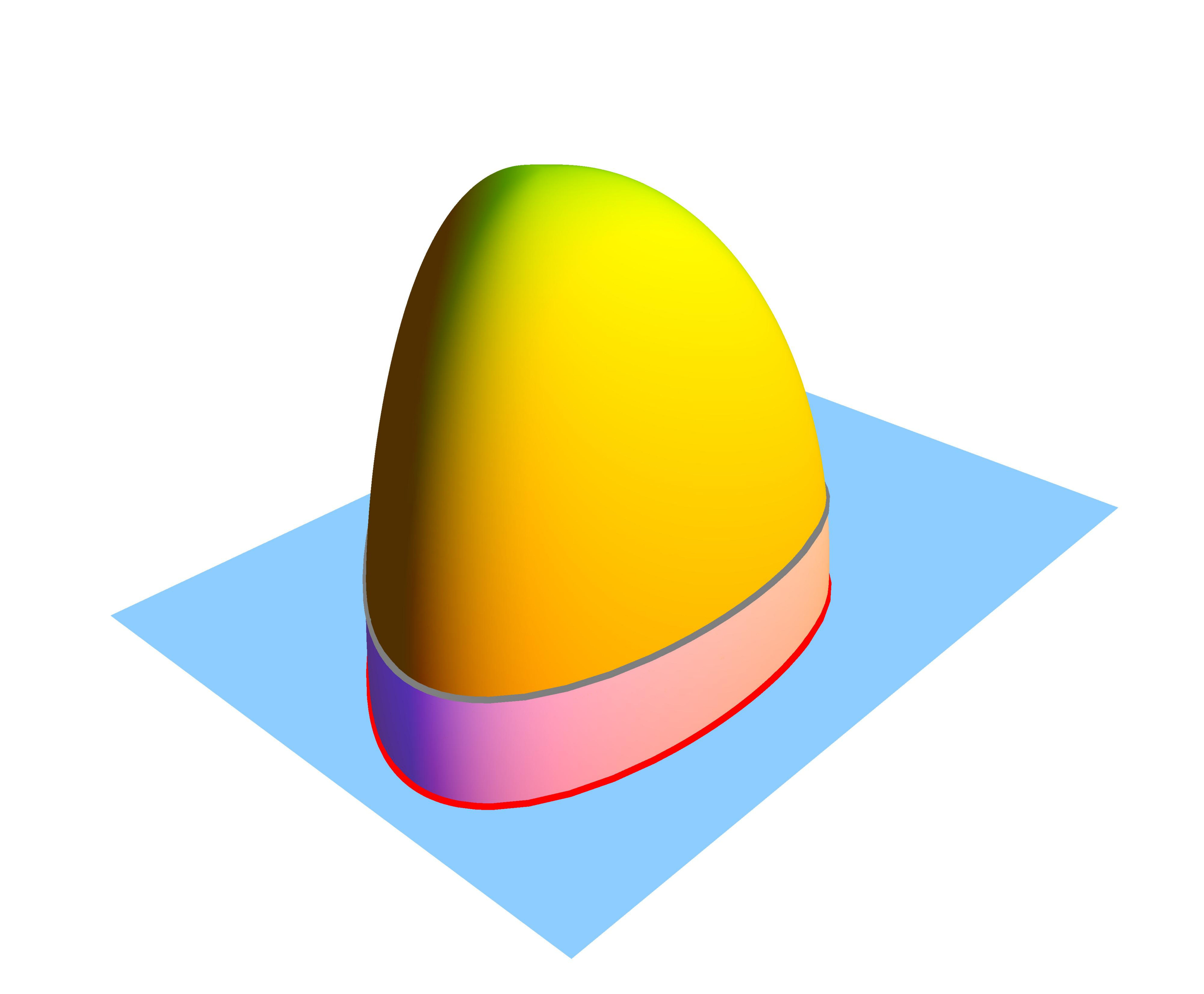}
   \label{fig3:subfig1}
 }
 \subfigure[]{
   \includegraphics[height=3cm] {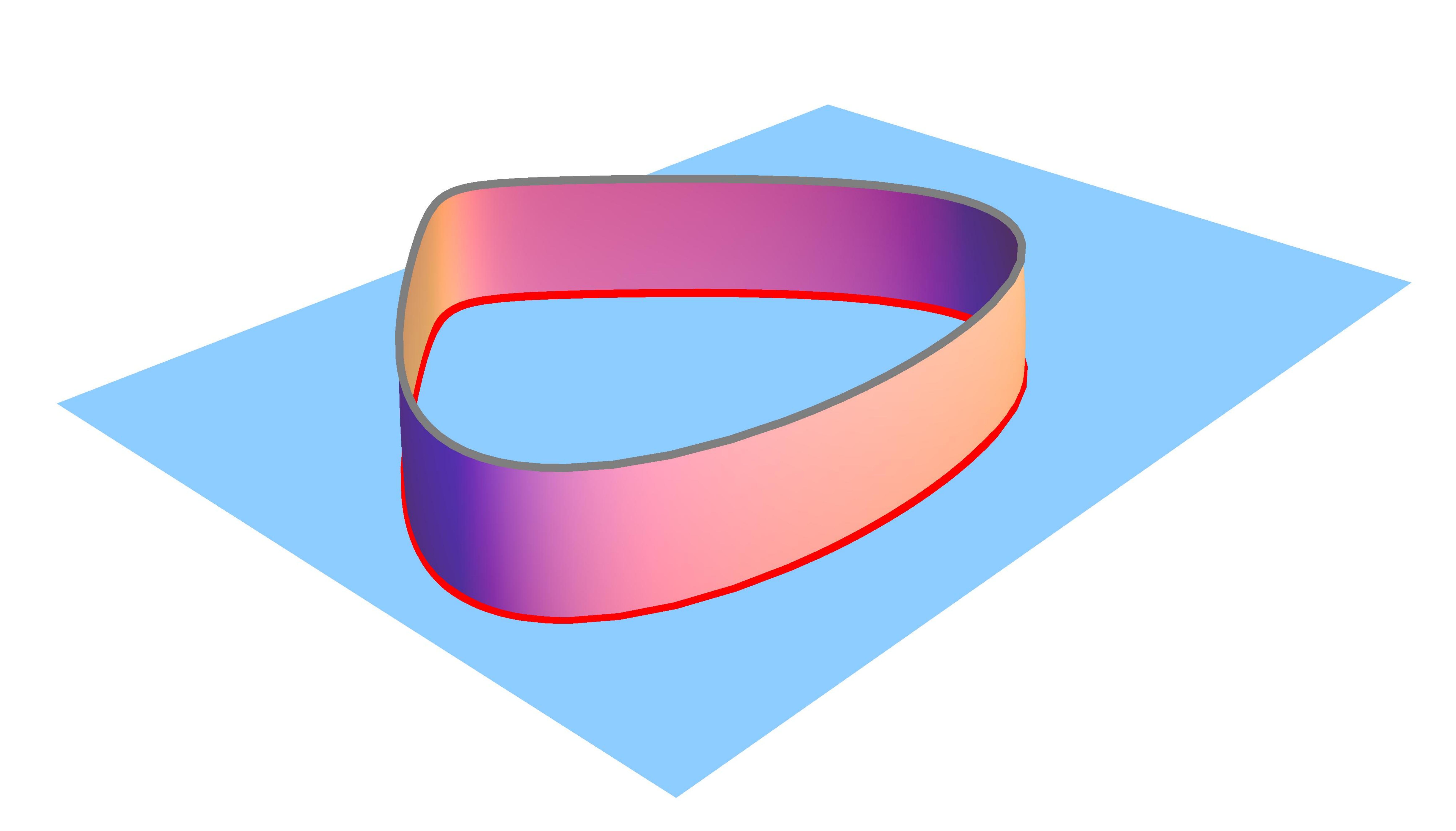}
   \label{fig3:subfig2}
 }
\caption{\label{}\small (a) A minimal surface in the Pilch-Warner geometry. The portion of the surface beyond the domain wall is shown in yellow. (b) The vertical part of the minimal surface described by the near-boundary asymptotics (\ref{nearbun}).}
\end{center}
\end{figure}

We would like to compute the minimal area for the surface that ends on a space-like contour $x^\mu (s)$ at $c=1$. The  surface will start off as a vertical wall, and then will shrink gradually some distance away from the boundary, fig.~\ref{fig3:subfig1}. In other words, near the boundary the solution for the minimal surface behaves as
\begin{equation}\label{nearbun}
 X^\mu (s,\tau )\simeq x^\mu (s),\qquad C(s,\tau )\simeq \tau .
\end{equation}
The bigger the contour  is, the farther in the bulk will the surface extend. For a very big contour the surface extends far beyond the domain wall, and the asymptotic solution (\ref{nearbun}) is a good approximation up to $c\sim c_0\gg 1$. Beyond the domain wall the metric takes a simple scaling form\footnote{A curious consequence of this geometric picture, observed in \cite{HoyosBadajoz:2010td}, is  that the theory becomes effectively five-dimensional in deep IR.} (\ref{farbeyond}). For a sufficiently big contour, the largest part of the surface will lie in this region (shown in yellow in fig.~\ref{fig3:subfig1}), reaching up to $c\sim L$, where $L$ is the size of the contour. But the area element for the metric (\ref{farbeyond}) scales as $1/c$, and the contribution of the beyond-the-domain region to the area actually goes to zero as $1/L$. This somewhat counterintuitive conclusion can be substantiated by an explicit calculation for the circle of a large radius \cite{Buchel:2013id}. The largest contribution to the area still comes from $c\sim 1$ even for very large contours. But for large loops the near-boundary solution (\ref{nearbun}) is a good approximation for $1\ll c\ll L$. For the sake of computing the area  we can thus replace the full surface by its cylindric truncation, fig.~\ref{fig3:subfig2}, with the induced metric
\begin{equation}
 ds_{\rm ind}^2=\frac{BM^2{x'} ^2ds^2}{\tau ^2-1}+\frac{d\tau ^2}{B\left(\tau ^2-1\right)^2}\,,
\end{equation}
for which we find
\begin{equation}
 A[x]=ML\int_{1+\frac{M^2}{2\Lambda _0^2}}^{\infty }
 \frac{d\tau }{\left(\tau ^2-1\right)^{\frac{3}{2}}}
 =\Lambda _0L-ML.
\end{equation}
Strictly speaking, the upper limit of integration should have been $c_0\gg 1$, but since the integral converges well, we have extended integration to infinity. The lower cutoff is chosen according to the relationship  (\ref{cr-}) between the radial coordinates in AdS and in the Pilch-Warner geometry. The subtraction of the linear divergence then is the same as in $AdS_5\times S^5$, which insures the continuity of the $M\rightarrow 0$ limit for small loops. After that we get $A_{\rm reg}=-ML$ and then perimeter law (\ref{perlaw}) follows from the area law (\ref{GeneralAreaLaw}) \cite{Buchel:2013id}.

We conclude that the free energy and the perimeter law for Wilson loops, computed holographically, reproduce the results obtained by direct path-integral calculation in field theory. Interestingly, the eigenvalue distribution itself can be also compared to supergravity, where it is determined by the D-brane probe analysis of the Pilch-Warner background. The distribution obtained this way appears to satisfy  the  Wigner law (\ref{WignerLaw}) with  $\mu =\sqrt{\lambda }M/2\pi $  \cite{Buchel:2000cn}. This is reproduced by the matrix-model result (\ref{strongmu})  in the decompactification limit, when  both $M$ and $\mu $ are rescaled by $R$ and $R$ is subsequently sent to infinity.

\section{Large-$N$ phase transitions: Super-QCD in Veneziano limit}

The decompactification limit of $R\rightarrow \infty $ can be regarded as a way to define the theory in flat space, as we discussed in sect.~\ref{sec:loc}. From this point of view, $R$ is an IR regulator that should be sent to infinity at the end of the calculation. The localization matrix model of $\mathcal{N}=2^*$ SYM simplifies dramatically in  this limit and  can be solved exactly for a finite interval of $\lambda $ \cite{Russo:2013qaa}. The solution terminates at the point of a fourth-order phase transition which happens at $\lambda _c\approx 35.4$ and is associated with new light states that appear in the spectrum \cite{Russo:2013qaa,Russo:2013kea}. Indeed, the mass of the hypermultiplet is not just $M$, but gets a contribution from the vacuum condensate (\ref{Phivev}), such that the mass squared of the component $\Phi ^{\rm hyper}_{ij}$ equals to $(a_i-a_j\pm M)^2$, and ranges from $(M-2\mu)^2$ to $(M+2\mu)^2$ as the eigenvalues $a_i$ and $a_j$ scan the interval $[-\mu ,\mu ]$. The correction due to the condensate is small compared to the bare mass only if $\mu \ll M$, which is true at weak coupling \cite{Russo:2013qaa}. But as $\mu $ grows with $\lambda $ and eventually reaches $M/2$,  a massless hypermultiplet
 contributes to the saddle-point, triggering the transition to the strong-coupling phase. As shown in \cite{Russo:2013qaa}, the theory undergoes secondary transitions each time the largest eigenvalue  satisfies the resonance condition $2\mu(\lambda _c^{(n)})=nM$. Since at strong coupling $\mu $ grows as $\sqrt{\lambda }M/2\pi $, there are infinitely many critical points  asymptotically approaching $\lambda _c^{(n)}\simeq \pi ^2n^2$. The resulting phase diagram looks like the one in fig.~\ref{fig:subfig2}. We are not going to discuss the phase structure of the $\mathcal{N}=2^*$ SYM in more detail, because of the technical complications (the details can be found in  \cite{Russo:2013qaa,Russo:2013kea}), and will instead consider a simpler model, where the same phenomena are under better analytical control.

The model that we are going to consider is  $\mathcal{N}=2$ super-QCD, by which we mean supersymmetric gauge theory with
$2N_f$ massive hypermultiplets of equal mass $M$. We shall assume that $N_f\leq N$, in which case the theory is asymptotically free.   The model 
interpolates between pure ${\cal N}=2$ SYM at $N_f=0$ and the mass deformation of superconformal YM at $N_f=N$.  We will study $\mathcal{N}=2$ SQCD in  
the Veneziano limit $N\rightarrow \infty $, $N_f\rightarrow \infty $ with $N_f/N$ fixed \cite{Veneziano:1976wm},  starting with the partition function on $S^4$ and subsequently taking the radius of the sphere to infinity.  

A neat way to define the partition function of $\mathcal{N}=2$ SQCD is to complement the theory with $2N-2N_f$ additional hypermultiplets of mass $M_{\rm 0}\gg M$. The theory with $2N$ hypermultiplets is a massive deformation of the $N_f=N$ superconformal theory, therefore it is finite. The partition function of this regularized version of $SQCD_{N_f}$ is
\be
Z_{N_f}^{\rm SQCD} =  \int d^{N-1}a\,\, 
\frac{\prod\limits_{i<j}^{} \left(a_i-a_j\right)^2 H^2(a_i-a_j)}{\prod\limits_{i}H(a_i+M)^{N_f} H(a_i-M) ^{N_f}H(a_i+M_0)^{N-N_f} H(a_i-M_0) ^{N-N_f}  }
\,{\rm e}^{- \frac{8\pi^2 N} {\lambda_0}\sum\limits_{i}  a_i ^2 }.
\ee
The heavy mass $M_0$ here acts as a UV cutoff. Taking $M_0\rightarrow \infty $, we expand
 $$
\ln H\left(a+M _0\right)+\ln H(a-M_0)\simeq
2\ln H(M_0)- 2\left(\ln M _0+\gamma +1\right)a^2
$$
where we used (\ref{functionKK}) and (\ref{Klog}). The large logarithm combines with the bare coupling into the dynamically generated scale
\begin{equation}
 \Lambda =M _0\,{\rm e}\,^{-\frac{4\pi ^2}{\lambda_0 \left(1-\zeta \right)}},
\end{equation}
where $\zeta $ is
the Veneziano parameter
\begin{equation}
\zeta =\frac{N_f}{N}\,.
\end{equation}
After the cutoff is removed, the partition function can  be written as
\be\label{NfQCD-partfunc}
Z^{\rm SQCD}_{N_f} =  \int d^{N-1}a\, \,\frac{\prod_{i<j}^{}\left(a_i-a_j\right)^2 H^2(a_i-a_j)}{\prod_{i}^{}H^{N_f}(a_i+M) H^{N_f}(a_i-M)  }
\,\,{\rm e}\,^{2(N-N_f )\left(\ln\Lambda +\gamma +1\right)\sum\limits_{i}  a_i ^2},
\ee
omitting an unimportant normalization constant.

As before, in the large-$N$ limit (with $\zeta $ fixed) the dynamics is described by the saddle-point equation:
\be
2\strokedint_{-\mu }^{\mu } dy \rho(y) \left(\frac{1}{x-y} -\KK(x-y)\right)
= -4\left(1-\zeta \right) \left(\ln \Lambda+\gamma +1\right)x - \zeta   \KK( x + M) - \zeta   \KK( x - M).
\label{fresa}
\ee
The  model depends on
two parameters $\Lambda$ and $M$, and greatly simplifies in the decompactification regime obtained by multiplying $\Lambda$, $M$, $\mu$,  $x$ and $y$  with $R$, which restores their canonical mass dimensions, and then sending $R$ to  infinity.  In this limit the arguments of the function $\KK $ are large and we can use the asymptotic formula (\ref{Klog}).
Differentiating the resulting equation in $x$ we get:
\be\label{auxiliary-SQCD}
2\int_{-\mu }^{\mu } dy\,\rho(y)\ln \frac{\left(x-y\right)^2}{\Lambda ^2}
= \zeta \ln\frac{\left(x^2-M^2\right)^2}{\Lambda ^4}\,.
\ee
Differentiating once more we obtain a singular integral equation that can be easily solved\footnote{The same equation appears in the matrix model for zero-dimensional open strings \cite{Kazakov:1989cq}, albeit with different  boundary conditions.}:
\be
\label{mfre}
2\strokedint_{-\mu }^{\mu } dy\,\,  \frac{\rho(y)}{x-y}
=   \frac{\zeta }{x+M} +\frac{\zeta }{x-M}\,.
\ee
The driving term in this equation has poles at $x=\pm M$ which may or may not lie within the eigenvalue distribution. The poles inside the distribution are due to resonances on massless hypermultiplets that appear when $a_i\pm M=0$. 
Depending on whether massless hypermultiplets appear or not, the model has two phases:
(1) the weak-coupling phase with $\mu <M$, in which all hypermultiplets are heavy, and (2) the strong-coupling phase at $\mu >M$, where light hypermultiplets appear in the spectrum. The solution to the saddle-point equations changes discontinuously when the pole at $x=M$ crosses the endpoint at $x=\mu $, and we need to consider the weak and strong coupling regimes separately.

\subsubsection*{Strong-coupling phase ($M<\mu $)} 

When $\mu >M$, the normalized eigenvalue density that solves the integral equation (\ref{mfre}) is  given by
\be
\label{noj}
\rho (x) = \frac{1-\zeta }{\pi \sqrt{\mu^2-x^2}} + \frac{\zeta }{2}\,\delta(x+M)+\frac{\zeta }{2}\,\delta(x-M).
\ee
We still need to fix $\mu $ in terms of $M$ and $\Lambda $. This is done by
 substituting the solution into (\ref{auxiliary-SQCD}). The latter equation is satisfied if 
\begin{equation}\label{mu=2Lambda}
 \mu =2\Lambda .
\end{equation}
The endpoint of the eigenvalue distribution turns out to be independent of the hypermultiplet mass.

The solution (\ref{noj}) is valid as long as $M<\mu $. When $M$ exceeds $\mu $, the delta-functions jump out of the interval $[-\mu ,\mu ]$ rendering the solution inconsistent.  As a result, the solution changes at the critical point and the system undergoes a transition to the weak-coupling regime. The phase transition thus happens at
\begin{equation}
 M_c=2\Lambda .
\end{equation}

\subsubsection*{Weak-coupling phase ($M>\mu $)}

 Assuming $\mu <M$ one finds the solution
\begin{equation}
 \rho (x)=\frac{1}{\pi \sqrt{\mu ^2-x^2}}\left(1-\zeta +\frac{\zeta M\sqrt{M^2-\mu ^2}}{M^2-x^2}
 \right).
\end{equation}
The integrated form of the saddle-point equation (\ref{auxiliary-SQCD}) then leads to
a transcendental equation for $\mu $:
\begin{equation}
2\left(1-\zeta \right)\ln\frac{\mu }{2M}+2\zeta \ln\frac{\mu }{M+\sqrt{M^2-\mu ^2}}=\left(1-\zeta \right)\ln\frac{\Lambda ^2}{M^2}\,.
\end{equation}
The solution can be conveniently expressed in a parametric form:
\begin{eqnarray}\label{muuu}
\mu &=&M\sqrt{1-u^2},
\\
 \left(\frac{2\Lambda }{M}\right)^{2-2\zeta }&=&\left(1+u\right)^{1-2\zeta }\left(1-u\right).
 \label{udefinition}
\end{eqnarray}

\subsubsection*{Critical behavior}

\begin{figure}[t]
\begin{center}
 \subfigure[]{
   \includegraphics[height=5cm] {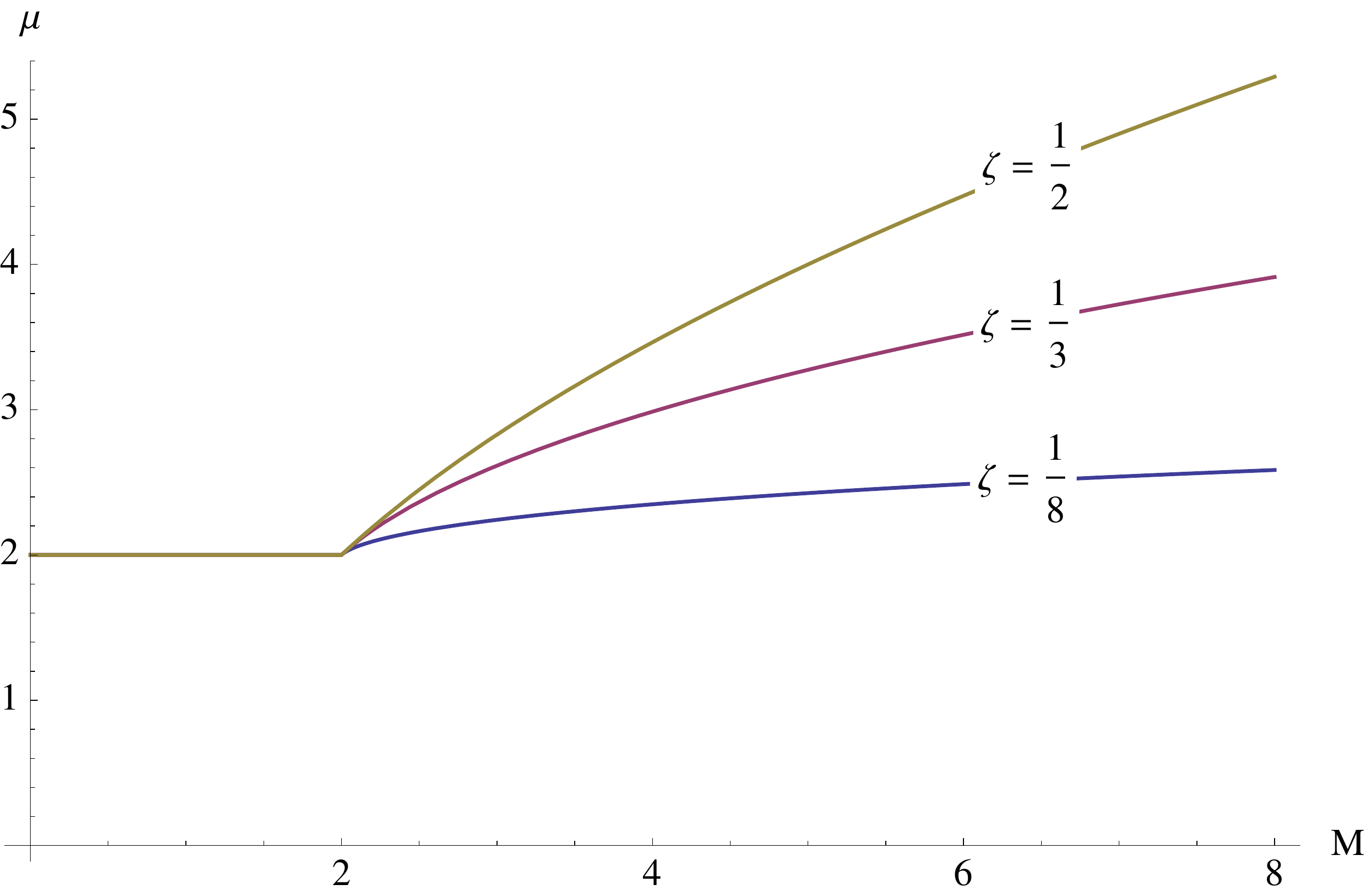}
   \label{fig4:subfig1}
 }
 \subfigure[]{
   \includegraphics[height=5cm] {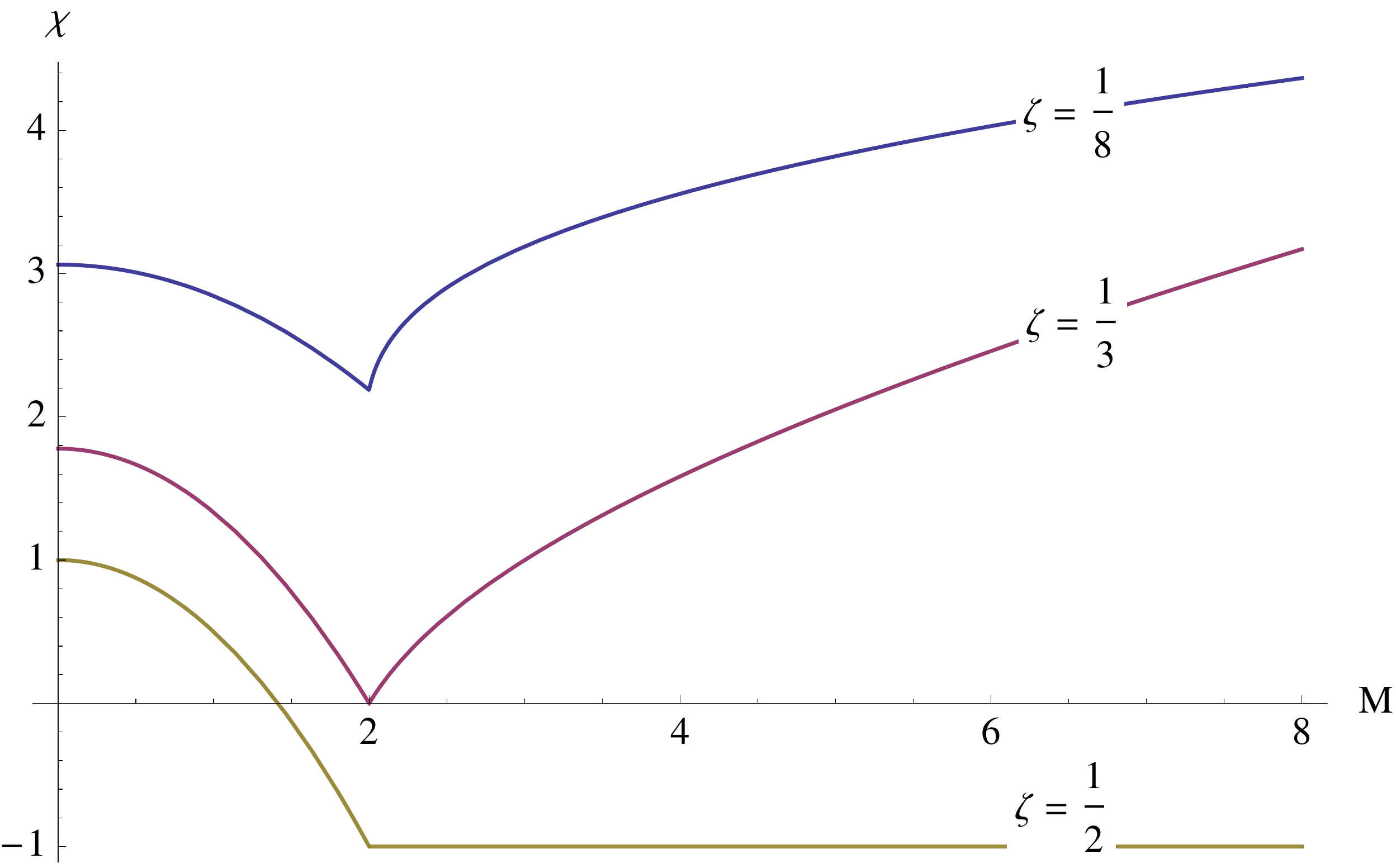}
   \label{fig4:subfig2}
 }
\caption{\label{mu-vs-M}\small The width of the eigenvalue distribution (left panel) and
susceptibility $\chi =-\partial ^2F/\partial \Lambda ^2$ (right panel)
 as  functions of the quark mass at fixed $\Lambda $ for different values of $\zeta $. The units on the $M$ and $\mu $ axes are normalized to $\Lambda $. The susceptibility is dimensionless.}
\end{center}
\end{figure}

From the solutions in the two phases we can calculate the Wilson loop and the free energy. The Wilson loop obeys the perimeter law, with the coefficient of proportionality equal to the largest eigenvalue of the matrix model:
\begin{equation}
 W(C)\simeq \,{\rm e}\,^{\mu L(C)}.
\end{equation}
As for the free energy, it is easier to compute its first derivative:
\begin{equation}
 \Lambda \,\frac{\partial F}{\partial \Lambda }=-2\left(1-\zeta \right)\left\langle x^2\right\rangle.
\end{equation}
We find:
\begin{equation}\label{defree}
 -\Lambda \,\frac{\partial F}{\partial \Lambda }=(1-\zeta )\times 
\begin{cases}
4\left(1-\zeta \right)\Lambda ^2+2\zeta M^2, & {\rm }M<2\Lambda 
\\
 \left(1-u\right)\left[1+\zeta +\left(1-\zeta \right)u\right]M^2,  & {\rm }M>2\Lambda 
\end{cases}
\end{equation}
where $u$ is the variable defined in (\ref{udefinition}). Comparing the two expressions at $M=2\Lambda $, one can show that $F$ is continuous across the transition together with its first and second derivatives, while the third derivative experiences a finite jump \cite{Russo:2013kea}. The transition thus is of the third order, as it usually happens in matrix models \cite{Gross:1980he,Wadia:2012fr}. 
The perimeter-law coefficient 
has discontinuous first derivative, as clear from fig.~\ref{mu-vs-M}.

\subsubsection*{Operator product expansion}

When the mass scale and the dimensional transmutation scale are widely separated, $M\gg \Lambda $,  hypermultiplets can be integrated out. 
What remains is pure gauge $\mathcal{N}=2$ SYM without matter. Taking into account the difference in the beta functions of pure $\mathcal{N}=2$ SYM and $\mathcal{N=2}$ SQCD, the dynamical scale of the low-energy effective field theory must be
\begin{equation}\label{Leff-SQCD}
 \Lambda _{\rm eff}=\Lambda ^{1-\zeta }M^\zeta .
\end{equation}
We may expect that the free energy in this regime has an OPE-type expansion of the form (\ref{genOPE}). On the other hand, the free energy can be calculated  from the solution of the matrix model.

Integrating (\ref{defree}), the free energy in the weak-coupling phase can be written explicitly in terms of the variable $u$ defined in (\ref{udefinition}):
\be
\label{freeuu}
F = - \frac{M^2}{2}  \left\{(1-\zeta ) (1-u) \left[1+ 5 \zeta +(1-\zeta ) u\right]+4  \zeta (1-2 \zeta )
   \ln \frac{1+u}{2}\right\}.
\ee
We have chosen the integration constant such that the free energy vanishes at $u=1$, which corresponds to the limit we are interested in, $\Lambda\ll M$. This expression can be now expanded in $1-u$. To facilitate this expansion it is convenient to introduce a new variable $v=(1-u)/2$. The equations (\ref{muuu}), (\ref{udefinition}), (\ref{freeuu}) then become
\begin{eqnarray}\label{mueff}
 &&\mu =2M\sqrt{v\left(1-v\right)}\\
 &&F=-2M^2\left\{
 \left(1-\zeta \right)\left[1+2\zeta -\left(1-\zeta \right)v\right]v
 +\zeta \left(1-2\zeta \right)\ln\left(1-v\right)
 \right\}
 \\ \label{veff}
 &&v\left(1-v\right)^{1-2\zeta }=\frac{\Lambda _{\rm eff}^2}{M^2}\,.
\end{eqnarray}
It is obvious from these expressions that the free energy has a power series expansion in $\Lambda _{\rm eff}^2/M^2$:
\begin{equation}
 F=\Lambda _{\rm eff}^2\left[
 -2+\zeta \,\frac{\Lambda _{\rm eff}^2}{M^2}+\frac{2}{3}\,\zeta \left(1-2\zeta \right)\frac{\Lambda _{\rm eff}^4}{M^4}
 +\frac{1}{6}\,\zeta \left(1-2\zeta \right)\left(5-8\zeta \right)\frac{\Lambda _{\rm eff}^6}{M^6}+\ldots 
 \right].
\end{equation}
Likewise,
\begin{equation}
 \mu ^2=4\Lambda _{\rm eff}^2\left[
 1-2\zeta \,\frac{\Lambda _{\rm eff}^2}{M^2}-3\zeta \left(1-2\zeta \right)\frac{\Lambda _{\rm eff}^4}{M^4}
 +\frac{4}{3}\,\zeta \left(1-2\zeta \right)\left(5-8\zeta \right)\frac{\Lambda _{\rm eff}^6}{M^6}+\ldots 
 \right].
\end{equation}
As we discussed in the introduction this expansion can be identified as arising from the OPE of the effective action induced be integrating out heavy hypermultiplets. The equations (\ref{mueff})--(\ref{veff}), which are exact, thus resum OPE to all orders.

Interestingly, in the particular case $\zeta =1/2$, the OPE truncates at the first order:
\begin{eqnarray}
 F_{\zeta =1/2}&=&-2\Lambda _{\rm eff}^2+\frac{1}{2}\,\,\frac{\Lambda _{\rm eff}^2}{M^2}\\
 \mu _{\zeta =1/2}^2 &=&4\Lambda _{\rm eff}^2-4\,\frac{\Lambda _{\rm eff}^4}{M^2}\,.
\end{eqnarray}
This suggests that superselection rules must exist in ${\cal N}=2$ SQCD with $N_f = N/2$  which set to zero the vevs of higher dimensional operators.

\section{Conclusions}

We have shown above how to solve for the large-$N$ master  field of ${\cal N}=2^*$ SYM,  and ${\cal N} = 2$ SQCD with $2N_f$ flavors using supersymmetric localization.  One of the important lessons  that one can draw from these calculations is the existence of quantum weak/strong coupling phase transitions, which 
seem to be generic features of  massive ${\cal N} = 2$ theories. There is a single third-order phase transition in SQCD, while ${\cal N} = 2^*$ theory exhibits an infinite number of large-$N$ phase transitions occurring as $\lambda$ is increased and accumulating towards $\lambda = \infty$ \cite{Russo:2013qaa,Russo:2013kea}.  At large $N$, the functional integral is dominated by a saddle point. Our calculation shows that, when the coupling overcomes a certain critical value (or several critical values, as in the case of ${\cal N} = 2^*$), this saddle-point  includes field configurations with extra massless hypermultiplets, thus producing discontinuities in vacuum expectation values of gauge invariant
observables. 

The free energy and the expectation values of large Wilson loops have only non-perturbative terms in their weak-coupling expansion. We have shown how to  compute the expansion coefficients for SQCD to any order  (the results for $\mathcal{N}=2^*$ theory can be found in \cite{Russo:2013qaa}).
Non-perturbative series of this type can be understood as  OPE  in the underlying field theory, arising due to large separation of scales.

The results of localization at strong coupling can be compared to prediction of the holographic duality. The results of explicit field-theory calculations perfectly agree with predictions of holography for the eigenvalue distribution, the vev of large Wilson loops \cite{Buchel:2013id} and the free energy on $S^4$  \cite{Bobev:2013cja}. We demonstrated this for the $\mathcal{N}=4$ and $\mathcal{N}=2^*$ SYM theories, holographic duals of which are explicitly known.

We conclude by mentioning a  number of  open problems. One important problem concerns additional checks  of holographic duality. In particular, 
the recent construction of the five-dimensional supergravity solution  dual to ${\cal N}=2^*$ compactified on $S^4$ \cite{Bobev:2013cja}
illustrates the way to construct {\it euclidean} gravity solutions representing supersymmetric gauge theories on  spaces of positive curvature.
This new type of solutions  would permit one to perform a number of new tests 
and thereby achieve a deeper understanding of gauge/gravity duality in non-conformal settings.

 At strong coupling, the phase transitions occur at $\sqrt{\lambda}\sim n\pi$, with positive integer $n\gg 1$. It would be extremely interesting to find a  string-theory interpretation  of these special values of $\lambda $. It is conceivable that some signs of the non-analyticity at  $\sqrt{\lambda}\sim n\pi$ could be manifested 
for semiclassical strings in the Pilch-Warner geometry \cite{Dimov:2003bh}.

In the case of pure  $\mathcal{N}=2$  SYM, it was  shown in \cite{Russo:2012ay} that in the decompactification limit localization reproduces the same
eigenvalue distribution that arises from the Seiberg-Witten solution \cite{Douglas:1995nw,Ferrari:2001mg}. This distribution arises in the limit of maximally degenerate curves.
It would be interesting to reproduce the results of ${\cal N}=2^*$ $SU(N)$ localization
from the corresponding Seiberg-Witten solution studied in \cite{Donagi:1995cf,DHoker:1997ha}.
It seems plausible that, like in pure  $\mathcal{N}=2$, there is a suitable limit that reproduces the same eigenvalue density found by localization and hence the same pattern of quantum phase transitions discussed here.

\subsection*{Acknowledgments}

The work of K.Z. was supported in part by People Programme (Marie Curie Actions) of the European Union's FP7 Programme under REA Grant Agreement No 317089. J.R. acknowledges support by MCYT Research
Grant No.  FPA 2010-20807.


\bibliographystyle{nb}

\end{document}